\documentstyle[preprint,aps]{revtex}
\begin{document}
\draft
\author{G. Lazarides\thanks{%
E-mail lazaride@vergina.eng.auth.gr}}
\address{Physics Division,School of Technology,\\
University of Thessaloniki,\\
Thessaloniki 540 06, Greece}
\author{N.D. Vlachos\thanks{%
E-mail vlachos@olymp.ccf.auth.gr. Work supported in part by E.U. grant
Science SC1-0729-92}}
\address{Dept. of Theoretical Physics\\
University of Thessaloniki\\
Thessaloniki 540 06, Greece}
\title{Multiinstanton Ladders in Baryon Number\\
Violating Processes }
\maketitle

\begin{abstract}
We estimate the contribution of a class of multiinstanton ladder graphs to
baryon and lepton number violating processes in the standard model. We find
that this contribution is negligible and does not alter the high energy
behavior of the leading semiclassical approximation.
\begin{center}(To appear in Phys.Rev. D) \end{center}
\end{abstract}

\newpage\

\section{Introduction}

\ Instanton-induced processes in the standard electroweak theory are known
to lead to baryon and lepton number violation. Although 't Hooft \cite{h1}
showed several years ago that such phenomena are utterly suppressed by the
factor $e^{-8\pi ^2/g^2}$ ($g$ is the $SU(2)_L$ gauge coupling constant),
several authors \cite{c2} explored the possibility that this exponential
suppression factor can be overcome at high energies by the phase space which
corresponds to multiparticle production. The key observation is that the $%
SU(2)_L$- instanton induces to leading semiclassical approximation effective
point-like interactions which involve all the fermionic left-handed $SU(2)_L$%
-doublets of the theory (four per generation) and any number of Higgs and
gauge bosons.

The inclusive cross section of the baryon and lepton number violating two
fermion scattering can then be calculated as the imaginary part of the
forward $2\rightarrow 2$ scattering amplitude depicted in fig.1 \cite{p3}.
As a result, this inclusive cross section appears to grow exponentially with
energy and can conceivably become unsuppressed at energies of the order of
the sphaleron \cite{m4} mass. This leading order behavior may, however, be
drastically altered by higher order corrections well before the energy
reaches the sphaleron mass and consequently, these phenomena may remain
unobservable at all energies. Several authors \cite{a5} actually, have
suggested that this could be the case if multiinstanton corrections were to
be taken into account. Corrections to the $2\rightarrow 2$ scattering
amplitude consisting of linear instanton-antiinstanton chains in alternating
order were considered, with particle exchange allowed only between
successive instantons and antiinstantons. Dorey and Mattis \cite{d6},
however, using the valley method \cite{k7}, pointed out that inclusion of
non-nearest neighbor instanton-antiinstanton as well as instanton-instanton
and antiinstanton-antiinstanton interactions could render these chain graphs
unimportant at the relevant energies. This, however, may not happen if
Dorey's result on the non linear $O(3)$ $\sigma -$model applies in the
realistic case too. The imaginary part of the chain graphs being
ultraviolately divergent,
requires the introduction of an appropriate cut-off. Finally, these graphs
do not include initial state corrections and thus, are not expected to alter
the high energy behavior of the leading semiclassical approximation \cite{k7}%
,\cite{m8}.

In this work we choose to deal with a class of ladder graphs shown in fig.2.
The imaginary part of these graphs turns out to be finite and can, in
principle, be unambiguously calculated. They can, in some sense, be thought
as including initial state corrections too since the incoming particles
enter in different instanton vertices. In addition, such ladder graphs are
known to dominate the high-energy behavior in ordinary field theories.We
should emphasize, however, that including the ladder graphs of fig.2 does
not solve the problem of initial state corrections. Indeed, the entire
picture of separated instanton-antiinstanton chains is an uncontrolled
approximation at energies where such chains actually become important.

\section{The leading semiclassical approximation.}

Consider the inclusive cross section, $\sigma _{inc}$, of the B and L
violating reaction
\begin{equation}
\label{eq1}q+q\rightarrow (3n_g-2)\overline{q}+n_g\overline{l}%
+\,any\,\#\,of\,Higgses,
\end{equation}
where $n_g$ is the number of fermion generations ($n_g\geq 1$), $q$ and $l$
represent quarks and leptons respectively and we have ignored for simplicity
the production of gauge particles. In the leading instanton approximation, $%
\sigma _{inc}$ can be determined by first calculating the forward $%
2\rightarrow 2$ scattering amplitude in Euclidean space as shown in fig.1.
Then, $\sigma _{inc}$ is given by the imaginary part acquired by this
amplitude after rotating the total incoming Euclidean particle momentum $%
p=p_1+p_2$ to Minkowski space $\ (p^2\rightarrow e^{-\imath \pi }\,p^2)$.
The expression corresponding to the Euclidean space forward scattering
amplitude is \cite{c2},\cite{p3}
\begin{equation}
\label{eq2}C\int_0^\infty d\rho ^2\,\rho ^{2\alpha }\,e^{-\pi ^2v^2\rho
^2}\,\int_0^\infty d\tilde \rho ^2\,\tilde \rho ^{2\alpha }\,e^{-\pi ^2v^2%
\tilde \rho ^2}\,\int d^4x\,e^{-\imath px}\,F(x^2,\rho ^2,\tilde \rho ^2).
\end{equation}
$C$ is a constant given by
\begin{equation}
\label{eq3}C\propto \frac 1{\pi ^2}\,(32\pi ^2)^{4n_g-1}\,\left( \frac{8\pi
^2}{g^2}\right) ^8\,\mu ^{\frac{43-8n_g}3}\,e^{-\frac{16\pi ^2}{g^2(\mu )}%
},\ \alpha =(28n_g+7)/12,
\end{equation}
$\rho $ and $\tilde \rho $ are the scale sizes of the instanton and the
antiinstanton, \thinspace $x_\mu $ is their Euclidean separation, $v=246GeV$
the electroweak breaking scale, g the gauge coupling and $\mu $ is the
renormalization point. The function $F(x^2,\rho ^2,\tilde \rho ^2)$, to
leading semiclassical order (or for $x^2\rightarrow \infty $), can be
written as
\begin{equation}
\label{eq4}F(x^2,\rho ^2,\tilde \rho ^2)\equiv F(x^2)=e^{\kappa /x^2}\,\frac
1{(x^2)^n},
\end{equation}
where $\kappa =\pi ^2\rho ^2\tilde \rho ^2v^2$ and $n=3(2n_g-1)\geq 3$. The
exponential factor in the formula above, corresponds to the Higgses in the
final state of reaction (1), whereas the second factor corresponds to the $%
4n_g-2$ fermions which are also being produced. Since we are interested in
the high energy behavior of $\sigma _{inc}$, we can assume throughout this
work that all fermions and Higgs bosons are effectively massless.

We will first consider the integral over the instanton-antiinstanton
separation
\begin{equation}
\label{eq5}I^0(p^2)\equiv \int d^4x\,e^{-\imath px}\,F(x^2)\quad .
\end{equation}
This integral converges at infinity for $n\geq 3$ , but diverges badly at $%
x=0$. This virulent ultraviolet divergence in Euclidean space is due to the
attractive nature of the Coulomb potential $V(x^2)=-\kappa /x^2$ resulting
from Higgs particle exchange between the instanton and the antiinstanton and
is an artifact of the leading semiclassical approximation. We, thus, define
a regularized integral
\begin{equation}
\label{eq6}I_\delta ^0(p^2)=\int_{x^2\geq \delta ^2}\,d^4x\,e^{-\imath
px}\,F(x^2)
\end{equation}
by removing from the range of integration a four-dimensional disc of finite
radius $\delta >0$ centered at the origin. Performing the angular
integrations, we obtain
\begin{equation}
\label{eq7}I_\delta ^0(p^2)=2\pi ^2\int_\delta ^\infty dr\,e^{\kappa
/r^2}\,r^{3-2n}\,G_{0\,2}^{1\,0}(\frac{p^2r^2}4|0,-1),
\end{equation}

\noindent where
\begin{equation}
\label{eq8}G_{0\,2}^{1\,0}(\frac{p^2r^2}4|0,-1)=\frac 1{2\pi \imath }%
\,\int_Ldz\left( \frac{p^2r^2}4\right) ^z\frac{\Gamma (-z)}{\Gamma (2+z)}
\end{equation}
is the well-known Meijer function. The contour $L$ is a loop starting and
ending at $+\infty $ and encircling the poles of $\Gamma (-z)$ once in the
negative direction. Since $p^2r^2/4$ is positive, one can show that the
contour $L$ can be distorted to become parallel to the imaginary axis and
lying in the strip $-1/2<Re(z)<0$. Then substituting eq.(8) in eq.(7) and
interchanging the order of integrations, we get

\begin{equation}
\label{eq9}I_\delta ^0(p^2)=\frac 1{2\pi \imath }\,\int_{c-\imath \infty
}^{c+\imath \infty }\,dz\,\left( \frac{p^2}4\right) ^z\,\frac{\Gamma (-z)}{%
\Gamma (2+z)}\,\pi ^2\,(-\kappa )^{z-n+2}\,\gamma (n-z-2,-\kappa /\delta
^2),
\end{equation}

$$
-1<c<0.
$$
Here, $\gamma (\alpha ,x)$ is the incomplete gamma function which can be
expressed as
\begin{equation}
\label{eq10}\gamma (\alpha ,x)=x^\alpha \,\Gamma (\alpha )\,\gamma
^{*}(\alpha ,x)\ ,
\end{equation}
with $\gamma ^{*}(\alpha ,x)$ being an analytic function of $\alpha $ and $x$%
.Eq.(9) then becomes
\begin{equation}
\label{eq11}I_\delta ^0(p^2)=\frac{\pi ^2\delta ^{4-2n}}{2\pi \imath }%
\,\int_{c-\imath \infty }^{c+\imath \infty }\,dz\left( \frac{p^2\delta ^2}4%
\right) ^z\,\frac{\Gamma (-z)\,\Gamma (n-2-z)}{\Gamma (2+z)}\,\gamma
^{*}(n-z-2,-\kappa /{\delta ^2}),
\end{equation}

$$
-1<c<0.
$$
Notice, that after interchanging the order of integrations, the range of $c$
can be extended.The imaginary part acquired by $I_\delta ^0(p^2)$ after
rotating $p^2$ to Minkowski space $(p^2\rightarrow e^{-\imath \pi }p^2)$,
comes from the $\ln p^2$ terms in the series expansion of the right-hand
side of eq.(11). These terms are produced by the double poles of the
integrand in eq.(11) at $z=m$,$~m=n-2,n-1,n,\ldots $ \nolinebreak. The
result is
$$
ImI_\delta ^0(e^{-\imath \pi }p^2)=\pi ^3\delta ^{4-2n}\sum_{m=n-2}^\infty
\left( \frac{p^2\delta ^2}4\right) ^m\,\frac{(-1)^{m-n+2}}{%
m!\,(m+1)!\,(m-n+2)!}\,\times
$$
\begin{equation}
\label{eq12}\gamma ^{*}(n-m-2,-\kappa /{\delta }^2).
\end{equation}
Using eq.(10), one can show that
\begin{equation}
\label{eq13}\gamma ^{*}(-\alpha ,x)=x^\alpha ,\quad \alpha =0,1,2,\ldots ,
\end{equation}
which implies
\begin{equation}
\label{eq14}ImI_\delta ^0(e^{-\imath \pi }p^2)=\pi ^3{\kappa }%
^{2-n}\,\sum_{m=n-2}^\infty \left( \frac{p^2\kappa }4\right) ^m\,\frac 1{%
m!\,(m+1)!\,(m-n+2)!}\,\cdot
\end{equation}
We now perform the integrals over the sizes of the instanton and the
antiinstanton to get the well-known semiclassical result.
\begin{equation}
\label{eq15}\sigma _{inc}^0=\frac{\pi ^3C}{(\pi ^2v^2)^{4+2\alpha -n}}%
\,\sum_{m=n-2}^\infty \left( \frac{p^2}{4\pi ^2v^2}\right) ^m\,\frac{[\Gamma
(\alpha +m-n+3)]^2}{m!\,(m+1)!\,(m-n+2)!}\,\ \cdot
\end{equation}
It is important to note that the $\delta $-dependence of the imaginary part
of $I_\delta ^0(e^{-\imath \pi }p^2)$ has completely disappeared in eq.(15).
Consequently, $\sigma _{inc}^0$ is $\delta $-independent for any $\delta >0$
and its exponential growth with energy results only from the boundary at
infinity of the Euclidean $x$-space in eq.(6).The virulent ultraviolate
divergence, as well as the contribution of any ''finite'' part of the
Euclidean $x$-space in eq.(6), do not seem to play any essential role.

\section{The ladder graphs.}

We will now turn to the calculation of the Euclidean space ladder graphs
shown in fig.2. The graphs, after rotating $p^2\rightarrow e^{-\imath \pi
}\,p^2$ and taking the imaginary part, constitute an important class of
multiinstanton corrections to the leading semiclassical approximation of $%
\sigma _{inc}$. The forward scattering amplitude which corresponds to the
ladder graph with $k+1$ rungs $(k=0,2,4,\ldots )$ is given by
\begin{equation}
\label{eq16}C^{k+1}\,\prod_{\imath =0}^k\,\prod_{\jmath =1}^k\,\int_0^\infty
\,d\rho _\imath ^2\,\rho _\imath ^{2\alpha }\,e^{-\pi ^2v^2\rho _\imath
^2}\,\int_0^\infty \,d\widetilde{\rho _\imath }^2\,\widetilde{\rho _\imath }%
^{2\alpha }\,e^{-\pi ^2v^2\widetilde{\rho }_\imath ^2}\times
\end{equation}
$$
\int \,d^4q_\jmath \,\frac 1{q_\jmath ^2}\,\int \,d^4x_\imath \,e^{-\imath
(q_{\imath }+q_{\imath +1})x_\imath }\,F(x_\imath ^2,\rho _\imath ^2,
\widetilde{\rho }_\imath ^2)\quad ,
$$
where $q_0=p_1$, $q_{k+1}=p_2$ and the four momenta $q_\jmath \,(\jmath
=1,2,\ldots ,k)$ are indicated in fig.2, $\rho _\imath \,,\widetilde{\rho
_\imath }\,\,\,(\imath =0,1,\ldots ,k)$ are the scale sizes of the $i$-th
instanton and the $i$-th antiinstanton respectively and $x_\imath $ is their
Euclidean separation.

We will first consider the integrals over the instanton-antiinstanton
separations
\begin{equation}
\label{eq17}I_\delta ^k(p^2)=\prod_{\jmath =1}^k\,\int \,d^4q_\jmath \,\frac
1{q_\jmath ^2}\,\,\prod_{\imath =0}^k\,\int\limits_{x_\imath ^2\geq \delta
^2}\,d^4x_\imath \,e^{-\imath (q_\imath +q_{\imath +1})}\,F(x_\imath ^2,\rho
_\imath ^2,\widetilde{\rho }_\imath ^2)\quad ,
\end{equation}
where the Euclidean space ultraviolet divergences are again regularized by
restricting the $x_i$-integrations to $x_\imath ^2\geq \delta ^2$. Repeating
the analysis of the previous section, we obtain\
$$
I_\delta ^k(p^2)=\frac{\pi ^{2(k+1)}\,\delta ^{(4-2n)(k+1)}}{(2\pi \,i)^{k+1}%
}\,\prod_{\imath =0}^k\,\int_{c_\imath -\imath \infty }^{c_\imath +\imath
\infty }\,dz_\imath \,\left( \frac{\delta ^2}4\right) ^{z_\imath }\,\frac{%
\Gamma (-z_\imath )\,\Gamma (n-2-z_\imath )}{\Gamma (2+z_\imath )}\times
$$
\begin{equation}
\label{eq18}\gamma ^{*}(n-z_\imath -2,-\frac{\kappa _\imath }{\delta ^2}%
)\,\,A_k(z_0,z_1,\ldots ,z_k)
\end{equation}
where $\kappa _\imath =\pi ^2\rho _\imath ^2\,\widetilde{\rho }_\imath
^2\,v^2$ and (see fig.2)
\begin{equation}
\label{eq19}A_k(z_0,z_1,\ldots ,z_k)=\prod_{\jmath =1}^k\,\int \,d^4q_\jmath
\,\frac 1{q_\jmath ^2}\,\prod_{\imath =0}^k\,(q_\imath +q_{\imath
+1})^{2z_\imath }\quad .
\end{equation}
The constants $c_\imath \ (\imath =0,1,\ldots ,k)$ which satisfy the
inequalities $-1<c_\imath <\nolinebreak0\,$(see eq.(11)) may and in fact
will have to be further restricted in eq.(18) so that $A_k(z_0,z_1,\ldots
,z_k)$ and the $z_\imath $ -integrals exist. For the moment we just assume
that there is some region of $z_\imath $ 's in which $A_k(z_0,z_1,\ldots
,z_k)$ exist and we restrict ourselves in this region. This assumption will
be proved to be correct a posteriori (see below). The $s=(p_1+p_2)^2$
dependence of $A_k$ at $p_1^2=p_2^2=0$ can be easily found from dimensional
arguments (there are no infrared divergences)\footnote{%
This semieuristic argument can be further corroborated by an explicit
although tedious calculation.}. We get

\begin{equation}
\label{eq20}A_k(z_0,z_1,\ldots ,z_k)=F_k(z_i)\,s^{k+\sum_i\,z_i}.
\end{equation}
It is easily seen that, for $p_1^2=p_2^2=0,$we also have
\begin{equation}
\label{eq21}p_2\cdot \frac \partial {\partial p_2}A_k=s\,\frac \partial {%
\partial s}A_k,
\end{equation}
and, thus,
\begin{equation}
\label{eq22}p_2\cdot \frac \partial {\partial p_2}A_k=(k+\sum_i\,z_i)A_k=z_k%
\,A_k-z_k\,A_k^{\,q_k}(z_0,\,z_1,\ldots ,z_k-1).
\end{equation}
\noindent $A_k^{\,q_k}$ denotes the expression $A_k$ with the $q_k$-
propagator ommited. Eq.(22 ) then gives
\begin{equation}
\label{eq23}A_k=-\frac{z_k}{k+\sum_{i\neq k}\,z_i}\,A_k^{\,q_k}(z_0,\,z_1,%
\ldots ,z_k-1)\ ,\ k+\sum_{i\neq k}\,z_i\neq 0\ \cdot
\end{equation}
The $q_k$-integration in $A_k^{\,q_k}$ can now be performed :
$$
\int d^4q_k\,[(q_k+p_2)^2]^{z_k-1}\,[(q_{k-1}+q_k)^2]^{z_{k-1}}=
$$

\begin{equation}
\label{eq24}\pi ^2\frac{\Gamma (-1-z_k-z_{k-1})}{\Gamma (1-z_k)\,\Gamma
(-z_{k-1})}\,{\rm B}(1+z_k,2+z_{k-1})\,[(q_{k-1}-p_2)^2]^{1+z_k+z_{k-1}}{\rm %
,}
\end{equation}

\noindent for $-1<Re(z_k)<1\,,\ -2<Re(z_{k-1})\,,\ Re(z_k)+Re(z_{k-1})<-1.$%
We then obtain the reccurence formula
$$
A_k(z_0,z_1,\ldots ,z_k)=\frac{\pi ^2}{\,[k+\sum_{j\neq k}z_j]}\,\frac{%
\,\Gamma (-1-z_k-z_{k-1})}{\,\Gamma (-z_k)\,\Gamma (-z_{k-1})}\times
$$

\begin{equation}
\label{eq25}\,B(1+z_k,\,2+z_{k-1})\,A_{k-1}(z_0,z_1,\ldots
,z_{k-2},\,1+z_k+z_{k-1})\ ,
\end{equation}
where $Re(z_k)<0$ and $A_k\quad (k=1,3,\ldots )$ is also defined by eq.(19)
but with $q_{k+1}=-p_2$ . Note that eq.(25) obviously holds for $%
k=1,3,5,\ldots $ too. Introducing the function
\begin{equation}
\label{eq26}D_k(z_0,z_1,\ldots ,z_k)=\prod_{\imath =0}^k\,\frac{\Gamma
(-z_i) }{\Gamma (2+z_\imath )}\,A_k(z_0,z_1,\ldots ,z_k)\quad k=0,1,2,\ldots
,
\end{equation}
the reccurence formula in eq.(25 ) takes the simple form
\begin{equation}
\label{eq27}D_k(z_0,z_1,\ldots ,z_k)=\frac{\pi ^2}{(1+z_k)(k+\sum_{\jmath
\neq k}z_\jmath )}\,\,D_{k-1}(z_0,z_1,\ldots ,z_{k-2},1+z_{k-1}+z_k)
\end{equation}
and can be easily solved to give
$$
D_k(z_0,z_1,\ldots ,z_k)=\pi ^{2k}\,\prod_{m=1}^k\,\frac 1{(m+z_0+\cdots
+z_{m-1})(k+1-m+z_m+\cdots +z_k)}
$$
\begin{equation}
\label{eq28}\times \frac{\Gamma (-k-\sum_{\imath =0}^k\,z_\imath )}{\Gamma
(2+k+\sum_{\imath =0}^k\,z_\imath )}\,s^{k+\sum_iz_\imath }.
\end{equation}
This formula holds provided that
$$
-2<Re(z_m)\quad m=0,1,\ldots ,k-1\ ;
$$
$$
-1<k-m+Re(z_m+\cdots +z_k)<0\quad m=1,2,\ldots ,k)\ ;
$$
$$
k+Re(z_0+\cdots +z_k)<0\ ;
$$
\begin{equation}
\label{eq29}m+z_0+\cdots +z_{m-1}\neq 0\quad m=1,2,\ldots ,k
\end{equation}
as can be easily deduced from the restrictions which follow eqs.(19), (24)
and (25). Substituting eq(28) in eq.(18) we obtain
$$
I_\delta ^k(s)=\frac{4^k\pi ^{2(2k+1)}}{(2\pi \imath )^{k+1}}\,\prod_{\imath
=0}^k\,\int_{c_\imath -\imath \infty }^{c_\imath +\imath \infty }\,dz_\imath
\,(\delta ^2)^{2-n+z_\imath }\,\Gamma (n-2-z_i)\,\gamma ^{*}(n-2-z_i,-\frac{%
\kappa _\imath }{\delta ^2})\times
$$
\begin{equation}
\label{eq30}\prod_{m=1}^k\frac 1{(m+z_0+\cdots +z_{m-1})(k+1-m+z_m+\cdots
+z_k)}\times
\end{equation}
$$
\frac{\Gamma (-k-\sum_{\imath =0}^kz_\imath )}{\Gamma (2+k+\sum_{\imath
=0}^kz_\imath )}\,\left( \frac s4\right) ^{k+\sum_\imath z_\imath }
$$
with $-1<c_\imath <0\quad (\imath =0,1,\ldots ,k)$ and $k+c_0+\cdots +c_k<0$.

The $z_\imath $- integrals can be evaluated by collapsing their contours to
the right and using residue calculus. Since we are only interested in the
imaginary part acquired by the amplitude when $s\rightarrow e^{-\imath \pi
}s $ , we only keep contributions proportional to $\ln s$. The relevant
contributions to the first $k\,$ $z_\imath $-integrals $(\imath =0,1,\ldots
,k-1)$ then come from the simple poles of the functions $\Gamma
(n-2-z_i)\,(\imath =0,1,\ldots ,k-1)$ whereas the $z_k$-integral gets
contributions from the double poles of the product $\,\Gamma (n-2-z_k)\Gamma
(-k-\sum_{\imath =0}^kz_\imath )$ where the first $k\,\ z_\imath $'s $%
(\imath =0,1,\ldots ,k-1)$ have already been substituted by integers. The
final result is%
$$
Im\,I_\delta ^k(se^{-\imath \pi })=4^k\pi ^{4k+3}\sum_{n-1\leq
l_0,l_1,\ldots ,l_k}\,\prod_{\imath =0}^k\,\frac{(\kappa _\imath )^{l_\imath
+1-n}}{(l_\imath +1-n)\,!}\times
$$
\begin{equation}
\label{eq31}\prod_{m=1}^k\,\frac 1{[l_0+\cdots +l_{m-1}][l_m+\cdots +l_k]}%
\times \frac{\left( s/4\right) ^{\sum_{\imath =0}^k\,l_\imath -1}}{%
[\sum_{\imath =0}^k\,l_\imath -1]\,!\,[\sum_{\imath =0}^k\,l_\imath ]\,!}
\end{equation}
and turns out to be again $\delta $-independent. Performing the $\rho
_\imath ^2\,,\widetilde{\rho }_\imath ^2$ -integrals in eq.(16) we finally
obtain the contribution of the ladder graph with $k+1$ rungs $%
(k=0,2,4,\ldots )$ to $\sigma _{inc}$ :%
$$
\sigma _{inc}^k=\frac 1{\pi \,s}\,\left[ \,4\pi ^4\,C\,(\pi
^2v^2)^{n-3-2\alpha }\right] ^{k+1}\,\,\sum_{n-1\leq l_0,l_1,\ldots
,l_k}\,\prod_{\imath =0}^k\frac{\left[ \Gamma (l_\imath +\alpha +2-n)\right]
^2}{\Gamma (l_\imath +2-n)}\,\times
$$
\begin{equation}
\label{eq32}\prod_{m=1}^k\,\frac 1{[l_0+\cdots +l_{m-1}][l_m+\cdots +l_k]}%
\times \frac{\left( s/4\pi ^2v^2\right) ^{\sum_{\imath =0}^k\,l_\imath }}{%
[\sum_{\imath =0}^k\,l_\imath -1]\,!\,[\sum_{\imath =0}^k\,l_\imath ]\,!}
\end{equation}

The multiple Series found for $\sigma _{inc}^k$, as it stands, looks very
complicated to be handled. We shall attempt to get an estimate by finding
suitable upper and lower bounds. In order to achieve this, we shall make
extensive use of the inequalities%
$$
l_0\,l_1\cdots l_k\leq \frac 1{(k+1)!}\,(l_0+l_1+\cdots +l_k)^{k+1}\quad ,
$$
\begin{equation}
\label{eq33}\frac 1{l_0\,l_1\cdots l_k}>\frac{(k+1)!}{(l_0+l_1+\cdots
+l_k)^{k+1}}\,\cdot
\end{equation}

Taking into account that $\alpha >2$ and defining $a=[\alpha ]+1$ we find
that
\begin{equation}
\label{eq34}\sigma _l^k<\sigma _{inc}^k<\sigma _u^k\,,
\end{equation}
where
$$
\sigma _l^k=\frac 1{\pi \,s}\,D^{k+1}\,2^k\,\,\,\sum_{n-1\leq l_0,l_1,\ldots
,l_k}\,\frac 1{\left[ \sum_{\imath =0}^kl_\imath \right] ^{n(k+1)-1}}\,\frac{%
\prod_{\imath =0}^k\Gamma (l_\imath )}{\left[ \Gamma (\sum_{\imath
=0}^kl_\imath )\right] ^2}\left[ \frac s{4\pi ^2v^2}\right] ^{\sum l_\imath
}\,,
$$
$$
\sigma _u^k=\frac 1{\pi \,s}\,D^{k+1}\frac 1{(n-1)^{2k}}\frac 1{(k+1)^{2a}}%
\frac 1{(k!)^{2(a+1)}}\left( \frac{a^a}{a!}\right) ^{2(k+1)}\times
$$
\begin{equation}
\label{eq35}\sum_{n-1\leq l_0,l_1,\ldots ,l_k}\,\left[ \sum_{\imath
=0}^kl_\imath \right] ^{2a(k+1)-1}\,\frac{\prod_{\imath =0}^k\Gamma
(l_\imath )}{\left[ \Gamma (\sum_{\imath =0}^kl_\imath )\right] ^2}\left[
\frac s{4\pi ^2v^2}\right] ^{\sum l_\imath },
\end{equation}
with $D=4\pi ^4\,C\,(\pi ^2v^2)^{n-3-2\alpha }\,.$

It is now clear that the next step must be the study of the multiple Series
\begin{equation}
\label{eq36}\Sigma _k(x)=\frac 1x\sum_{l_\imath =1}^\infty \frac{\Gamma
(l_0)\cdots \Gamma (l_k)}{\left[ \Gamma (l_0+l_1+\cdots +l_k)\right] ^2}%
\,x^{l_0+l_1+\cdots +l_k}\,,\quad x=\frac s{4\pi ^2v^2}\,\,\cdot
\end{equation}
Then, $\sigma _l^k$ and $\sigma _u^k$ could be recovered by integrating or
differentiating $\Sigma _k(x)$ with respect to $\ln (x)$ a suitable number
of times. The fact that the $l_\imath $-summations in the definition of $%
\sigma _l^k$ and $\sigma _u^k$ start at $n-1$ cannot change the results in
any fundamental way.

The Laplace transform of $\Sigma _k(x)$ is%
$$
S_k(t)=\int_0^\infty dx\,e^{-xt}\Sigma _k(x)=
$$
\begin{equation}
\label{eq37}\sum_{l_\imath =1}^\infty \frac{\Gamma (l_0)\,\Gamma (l_1)\cdots
\Gamma (l_k)}{\Gamma (l_0+l_1+\cdots +l_k)}\,t^{-(l_0+l_1+\cdots +l_k)}
\end{equation}
We can now use the integral representation for the generalized beta function
\begin{equation}
\label{eq38}\frac{\Gamma (l_0)\,\Gamma (l_1)\cdots \Gamma (l_k)}{\Gamma
(l_0+l_1+\cdots +l_k)}=\int_0^1\,\prod_{\imath =0}^k\,d\alpha _{\imath
\;}\prod_{\jmath =0}^k\,\alpha _\jmath ^{l_\jmath -1}\,\delta (1-
{\textstyle \sum }\alpha_\imath )
\end{equation}
to get
\begin{equation}
\label{eq39}S_k(t)=\int_0^1\,\frac{\prod d\alpha _\imath }{\prod \alpha
_\imath }\delta (1-{\textstyle \sum } \alpha _\imath
)\,\sum_{l_\imath =1}^\infty \left[ \frac{\alpha _0}t\right] ^{l_0}\cdots
\left[ \frac{\alpha _k}t\right] ^{l_k}.
\end{equation}
The summations are now decoupled and can be readily performed provided that $%
\mid \alpha _\imath /t\mid <1.\quad $ We obtain
\begin{equation}
\label{eq40} S_k(t)=\int_0^1 \frac
{ \prod d\alpha _i\,\delta (1-{\textstyle \sum } \alpha _\imath ) }
{(t-\alpha _0)\,(t-\alpha _1)\cdots (t-\alpha _k) }
\quad \cdot
\end{equation}
The inverse Laplace transform of the expression above can be found and the
answer is
\begin{equation}
\label{eq41} S_k(x)=\int_0^1 \prod d\alpha _i\,\delta (1-{\textstyle \sum }
\alpha _\imath) \sum_{m=0}^k\,\frac {e^{\alpha _mx}}{P_m(\alpha _m)}
\end{equation}
where $P_m(y)$ is a polynomial given by
\begin{equation}
\label{eq42}P_m(y)=\,\frac{\prod_{\imath =0}^k(y-\alpha _\imath )}{y-\alpha
_m}\quad \cdot
\end{equation}
Now the $\alpha _\imath -$ integrations can be performed and we end up with
the following recursive formula
\begin{equation}
\label{eq43}\Sigma _k(x)=e^{\frac xk}\int_0^xdz\,e^{-\frac zk}\int_0^1d\rho
\,\Sigma _{k-1}(\rho \,z)\,\quad \cdot
\end{equation}
This integral equation can be transformed into an integrodifferential
equation
\begin{equation}
\label{eq44}\frac d{dx}\Sigma _k(x)=\frac 1k\,\Sigma _k(x)+\frac 1x%
\,\int_0^x\,\Sigma _{k-1}(z)\,dz\,\,
\end{equation}
or a differential equation
\begin{equation}
\label{eq45}x\,\frac{d^2}{dx^2}\,f_k(x)\,+\,\left[ \left( 2-\frac 1k\right)
x+1\right] \,\frac d{dx}f_k(x)\,+\,\left( 1-\frac 1k\right) \left(
x+1\right) \,f_k(x)=\,f_{k-1}(x)\,,
\end{equation}
where
\begin{equation}
\label{eq46}f_k(x)=e^{-x}\,\Sigma _k(x)\,\cdot
\end{equation}
When $x$ is large, the differential equation reduces to
\begin{equation}
\label{eq47}\,\frac{d^2}{dx^2}\,f_k(x)\,+\,\left[ 2-\frac 1k\right] \,\frac d%
{dx}f_k(x)\,+\,\left[ 1-\frac 1k\right] \,f_k(x)=\frac{\,f_{k-1}(x)}x\quad
\cdot
\end{equation}
One particular integral of this equation is
\begin{equation}
\label{eq48}f_k(x)=\frac k{k-1}\,\frac 1{1+\frac d{dx}}\,\frac 1{1+\frac k{%
k-1}\frac d{dx}}\,\frac{\,f_{k-1}(x)}x
\end{equation}
with $\,f_1(x)=1\,.$

\noindent It is now obvious that the leading order solution of eq.(45 ) can
be written as
\begin{equation}
\label{eq49}f_k(x)=\frac k{x^{k-1}}
\end{equation}
leading to
\begin{equation}
\label{eq50}\Sigma _k(x)=k\ \frac{e^x}{x^{k-1}}\,\cdot
\end{equation}
We have already pointed out that the upper and lower bounds for $\sigma
_{inc}^k$ which were defined in eq.( 35) can be related to derivatives or
integrals of $\Sigma _k(x)$ with respect to$\ \ln x$. Such operations,
however, cannot modify the form of $\Sigma _k(x)$ in an essential way since
the exponential growth cannot be affected . We do expect a change in the
leading power of $x$ and of course the constant coefficient will be
different. We conclude that $\sigma _l^k\sim c_l^k\,e^x\,x^{-m_1},\,\sigma
_u^k\sim c_u^k\,e^x\,x^{-m_2}$ where the constants $c_l^k$, $c_u^k$ , $m_1$
, $m_2$ can in principle be calculated. The cross section $\sigma _{inc}^k$
being bound between two exponentials, can only behave exponentially, possibly
modified by an asymptotic Series of inverse powers of $s$ . Taking into
account that $\sigma _{inc}^k$ for $k>1$ is highly supressed by the small $%
D^{k+1}$ factor containing the instanton 't Hooft factor, we deduce that the
contribution of all ladder graphs for $k>1$ is negligible and cannot alter
the high energy behavior of the leading order result. In particular they do
not affect the possible validity of the ZMS picture, that is based on the
instanton-antiinstanton chain graphs only.

This result is not totally unexpected. It is known that to each shaded blob
of fig.2, which represents exchange of any number of bosons and $4n_g-2$
fermions between an instanton and an intiinstanton, corresponds an
exponentially growing factor, while, to each instanton or antiinstanton, an
exponentially small 't Hooft factor. Since the number of instanton or
antiinstanton vertices outnumbers the number of the multiparticle-exchange
blobs by a factor of two, we can expect that the contribution of such
ladders is supressed compared to the leading semiclassical result. Moreover,
the fact that not all the momentum flows through each rung makes at the end
the ladder graphs grow only exponentially with $s$. This is in contrast to
the case of a linear chain where all momentum flows between the instanton
and the antiinstanton, creating thus an exponential growth that can
counterbalance the suppression effect of the 't Hooft factors.

We thank Q. Shafi and C. Bachas for collaborating in early stages of this
work.\

\newpage
\begin{center}
{\bf FIGURE CAPTIONS}
\end{center}

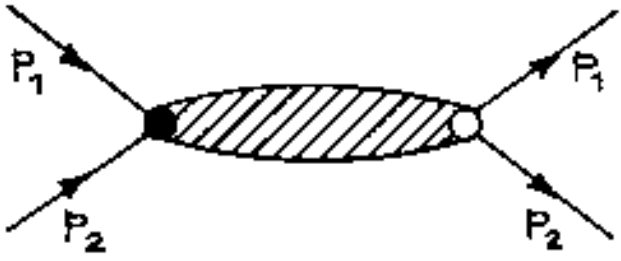
{\bf Fig.1} . The graph which correponds to the leading semiclassical
approximation to the forward $2\rightarrow 2$ scattering amplitude. Single
lines represent fermions, the filled and blank circles represent instanton
and antiinstanton vertices respectively, whereas the shaded blob represents
the exchange of $4n_g-2$ fermions and any number of Higgs bosons. $n_g$ is the
number of fermion generations.\\\\

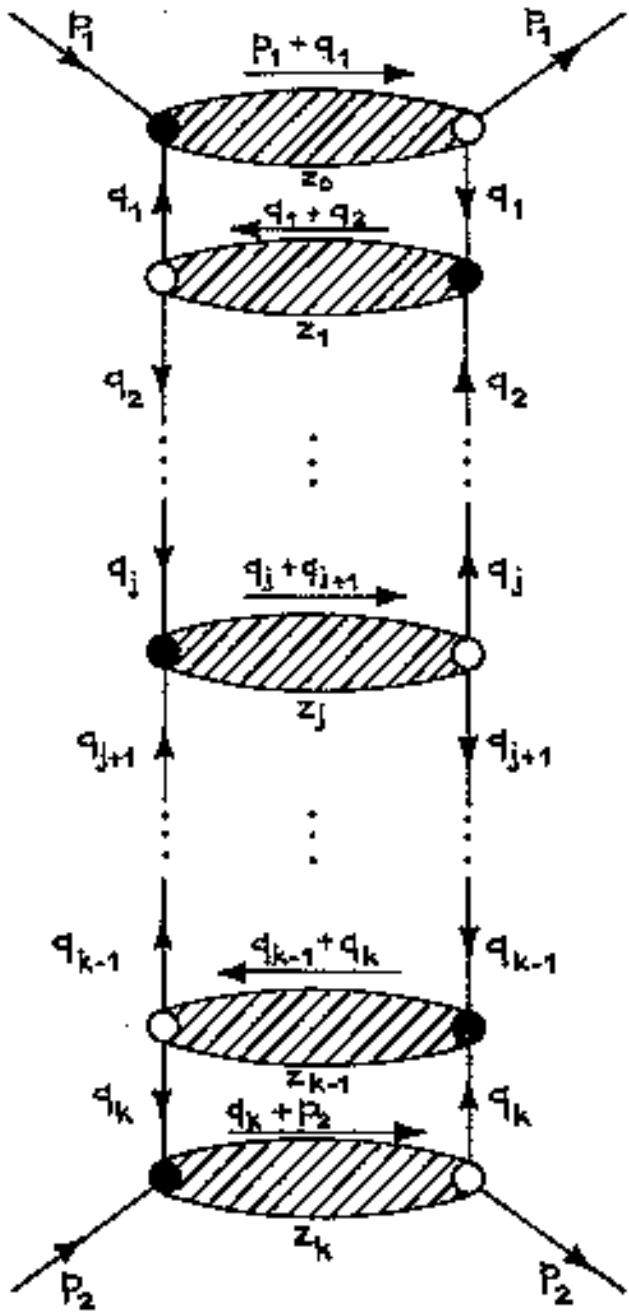
{\bf Fig.2} . The ladder graph with $k+1$ rungs $(k=0,2,4\ldots )$ .
Notation as in fig. 1. The complex parameters $z_\imath \,(\imath
=0,1,\ldots ,k)$ which appear in eq.(18) are also indicated.

\end{document}